\documentclass{ifacconf}

\usepackage{graphics} 
\usepackage{epsfig} 
\usepackage{ifpdf}
\ifpdf
  \usepackage{epstopdf}
\fi
\usepackage{mathptmx} 
\usepackage{times} 
\usepackage{amsmath} 
\usepackage{amssymb}  
\usepackage{floatrow}
\usepackage{multirow}
\usepackage{xcolor}
\usepackage{upgreek}
\usepackage{ mathrsfs }
\usepackage{lipsum}
\usepackage[parfill]{parskip}
\usepackage{tabularx, calc}
\usepackage{cite}
\usepackage{graphicx}      
\usepackage{natbib}        

\newtheorem{definition}{\indent Definition}

\DeclareMathAlphabet{\mathcal}{OMS}{cmsy}{b}{n}
\DeclareMathAlphabet{\mathcal}{OMS}{cmsy}{m}{n}

\begin{document}
\begin{frontmatter}

\title{Hybrid Filtering for a Class of Quantum Systems with Classical Disturbances\thanksref{footnoteinfo}}

\thanks[footnoteinfo]{This work was supported by the Australian Research Council's Discovery Projects funding scheme under Project DP130101658 and Laureate Fellowship FL110100020.}

\author[1]{Qi Yu}
\author[1]{Daoyi Dong}
\author[1]{Ian R. Petersen}
\author[1,2]{Qing Gao}

\address[1]{School of Engineering and Information Technology, University of New South Wales, Canberra, ACT 2600, Australia (e-mail: qi.yu@student.adfa.edu.au; i.r.petersen@gmail.com; daoyidong@gmail.com)}
\address[2]{Department of Mechanical and Biomedical Engineering, City University of Hong Kong, Hong Kong SAR, China (e-mail: qing.gao.chance@gmail.com)}

\begin{abstract}                
A filtering problem for a class of quantum systems disturbed by a classical stochastic process is investigated in this paper. The classical disturbance process, which is assumed to be described by a linear stochastic differential equation, is modeled by a quantum cavity model. Then the hybrid quantum-classical system is described by a combined quantum system consisting of two quantum cavity subsystems. Quantum filtering theory and a quantum extended Kalman filter method are employed to estimate the states of the combined quantum system. An estimate of the classical stochastic process is derived from the estimate of the combined quantum system. The effectiveness and performance of the proposed methods are illustrated by numerical results.
\end{abstract}

\begin{keyword}
quantum filtering, hybrid quantum-classical system, quantum extended Kalman filter. 
\end{keyword}

\end{frontmatter}


\section{Introduction}
Characterizing unknown quantum states have been a fundamental task in quantum computation, quantum metrology and quantum control. To estimate an unknown static quantum state, state tomography methods such as maximum likelihood estimation (\cite{book1}), Bayesian mean estimation (\cite{book1}) and linear regression estimation (\cite{qi:2013,hou:2016}) have been developed. For estimating a dynamic quantum state, a quantum filtering theory has been developed (\cite{bouten2007introduction,bouten2009introduction}). Quantum filtering theory was introduced by Belavkin in the 1980's as documented in a series of articles (\cite{belavkin1991continuous}). The basic premise is to build a non-commutative counterpart for classical probability theory so that approaches to deriving the classical filtering equation can be adapted to quantum dynamical systems. The main difference between this theory and classical filtering theory is that non-commutative observables in quantum systems cannot be jointly represented on a single classical probability space. Quantum filtering theory enables us to optimally estimate the quantum system state using non-demolition measurements. It plays a crucial role in many areas such as quantum control (\cite{van2005feedback}, \cite{armen2002adaptive}). Recently, quantum filtering theory has been successfully applied in experimental designs such as trapped ions \citep{hume2007high}, cavity QED systems \citep{sayrin2011real}, and optomechanical systems \citep{wieczorek2015optimal}.  
 In practice, physical quantum systems are unavoidably affected by classical signals \citep{wang2016fault,ralph2011frequency}, and a number of researchers are becoming interested in the filtering problem for `hybrid' quantum-classical systems where the quantum systems are subject to a classical process. Relevant results can be found in e.g., Tsang's work on quantum smoothing (\cite{tsang2009optimal,tsang2009time}) where a concept of hybrid quantum-classical density operator was used as the main technical tool.   Recently, \cite{gao2016fault,gao2016fault2} developed a quantum-classical Bayesian inference approach to solve fault tolerant quantum filtering and fault detection problems for a class of quantum optical systems subject to stochastic faults. \\
\\
In this paper, we extend the previous work \citep{gao2016fault} to the case that the disturbance process has a continuous value space and our main goal is to estimate both the quantum state and the classical process using non-demolition quantum measurements. We consider a system-probe model with a time-varying Hamiltonian that depends on a classical stochastic process. This hybrid quantum-classical stochastic system is analyzed by building a quantum analog of the classical stochastic process; see also \citep{wang2013quantum}. The idea of using an aritificial quantum system to model noise has been considered before \citep{xue2016feedback,xue2016quantum,xue2015quantum}. However, the author only consider the disturbance to be quantum noise.  Then, in our case, quantum filtering theory can be utilized to investigate the filtering problem. The estimation tasks are accomplished by using a quantum extended Kalman filter (QEKF) approach. \\
\\
The structure of this paper is as follows. In Section \ref{Sec2}, we briefly introduce quantum probability theory and quantum filtering theory.  Section \ref{Sec3} is devoted to the modeling of the classical signal using a quantum cavity model. A stochastic master equation (SME) is then obtained to solve the filtering problem. A QEKF approach is also employed to estimate both the quantum state and the classical process in Section 4. In Section \ref{Secsim}, we present a numerical example to demonstrate the performance and also compare the QEKF algorithm with the SME method. Section \ref{Conclusion} concludes this paper.\\
\\
Notation: $A_{m\times n}$ denotes an $m$ row and $n$ column matrix;
$A^\dagger$ denotes conjugate and transpose of $A$; $A^\top$ is the transpose of $A$; $A^*$ is the conjugate of $A$; $\text{Tr}(A)$ is the trace of $A$; $X$ is used to denote any operators and $x$ is a vector of those operators; $\rho$ is a density operator representing a quantum state; $\hat{a}$ is the estimate of $a$; $i$ means the imaginary unit, i.e., $i=\sqrt{-1}$.

\section{Preliminaries} \label{Sec2}

\subsection{Quantum probability theory}
We briefly present a preliminary discussion on quantum probability theory. For a detailed treatment, one can refer to the paper \citep{bouten2007introduction}. Denote the Hilbert space under consideration as $\mathscr{H}$. The system observables, which represent the physical properties of the system, are represented by self-adjoint operators on $\mathscr{H}$. The quantum state, which provides the status of a physical system, is specified by a density operator $\rho \in \mathcal{S(\mathscr{H})}$, where $\mathcal{S}$ is the class of unity trace operators on the associated Hilbert space \citep{emzir2016quantum}. In this paper, the evolution of the quantum system is mostly described under the Heisenberg picture. That means, any system observable evolves with time as $A(t)= U(t)^\dag A U(t)$ while the density operator $\rho$ remains unchanged.  Then any simple measurement of $A(t)$ yields values within the spectrum of $A(t)$ with a certain probability distribution and the expectation of the measurement is given by $\langle A(t) \rangle=\text{Tr}[\rho A(t)]$ \citep{bouten2007introduction}.
The key point of the quantum probability formalism is that any single realization of a quantum measurement corresponds to a particular choice of a commutative $*$-algebra of observables and any commutative $*$-algebra is equivalent to a classical (Kolmogorov) probability space \citep{bouten2007introduction}.\\
\\
For the finite-dimensional case, the set $\text{spec}(A)= \{a_j\}$ of eigenvalues of $A$ is called the spectrum of $A$, and $A$ can be written as
\begin{equation}\label{eq0.01}
A = \sum_{a\in \text{spec}(A)}aP_a,
\end{equation}
 where $P_a$ is the projection operator of $A$. The following theorem has been presented in \citep{bouten2007introduction}.
\begin{thm}\label{finitespectral}
	\citep{bouten2007introduction} (spectral theorem, finite-dimensional case). Let $\mathscr{A}$ be a commutative *-algebra of operators on a finite-dimensional Hilbert space, and let $\mathbb{P}$ 
	be a state on $\mathscr{A}$. Then there is a probability space $(\Omega,\mathcal{F},P)$ and a map $\iota$ from $\mathscr{A}$ onto the set of measurable functions on $\Omega$ that is a $*$-isomorphism; i.e., a linear bijection with $\iota (AB) = \iota(A) \iota(B)$ (pointwise) and $\iota(A^*)=\iota(A)^*$, and moreover $\mathbb{P}(A)=E_{\bold{P}}(\iota(A))$.
\end{thm}

For the infinite-dimensional case, a system operator can be expressed in terms of its spectral measure by
\begin{equation}\label{eq0.02}
A = \int_R \lambda P_A (d\lambda).
\end{equation}
The corresponding spectral theorem	for infinite-dimensional case is stated as follows \citep{bouten2007introduction}:
\begin{thm}\label{infinitespectral}
	\citep{bouten2007introduction} (Spectral Theorem). Let $\mathscr{C}$ be a commutative von Neumann algebra. Then there is a measure space $(\Omega, \mathcal{F},\mu)$ and a $*$-isomorphism $\iota$ from $\mathscr{C}$ to $L^\infty(\Omega, \mathcal{F},\mu)$, the algebra of bounded measurable complex functions on $\Omega$ up to $\mu$ - a.s. equivalence. Moreover, a normal state $\mathbb{P}$ on $\mathscr{C}$ defines a probability measure $\bold{P}$, which is absolutely continuous with respect to $\mu$ such that $\mathbb{P}(C)=E_{\bold{P}}(\iota(C))$ for all $C \in \mathscr{C}$.
\end{thm}
The spectral theorem above allows us to treat any set of commutative observables as a set of classical random variables defined on a single classical probability space. In other words, any quantum probabilistic concept can be directly extended to its classical counterpart. Therefore, classical statistical analysis methods can be applied directly in analyzing quantum systems. The following concept of quantum conditional expectation is defined in a similar way to classical conditional expectation and is very useful in quantum filtering theory \citep{bouten2007introduction}.
\begin{definition}
	\citep{bouten2007introduction}  (conditional expectation). Let $(\mathscr{N},\mathbb{P})$ be a quantum probability space and let $\mathscr{A}\subset \mathscr{N}$ be a commutative von Neumann subalgebra. Then the map $\mathbb{P}(.|\mathscr{A}): \mathscr{A}^\prime \rightarrow \mathscr{A}$ is called (a version of) the conditional expectation from $\mathscr{A}^\prime$ on to $\mathscr{A}$ if $\mathbb{P}(\mathbb{P}(B|\mathscr{A})A)=\mathbb{P}(BA)$ for all $A\in \mathscr{A}, B\in \mathscr{A}^{\prime}$ .
\end{definition}
 The $\mathscr{A}^\prime$ here is used to denote the commutant of $\mathscr{A}$. $\mathbb{P}(B|\mathscr{A})$ is the projection of $B$ onto the algebra $\mathscr{A}$ and represents the maximum information of $B$ that can be extracted from the observation $\mathscr{A}$.

\subsection{Quantum filtering theory}

We use quantum stochastic differential equations (QSDEs) to describe the dynamics of an open quantum system with driving noises. Three fundamental noise processes are described using the annihilation process $A_t$, the creation process $A_t^*$ and the Poisson (conservation) process $\Lambda_t$. The quantum $It\hat{o}$ integral is defined for the calculation of a quantum stochastic integral. \\
\\
With a corresponding conditional quantum expectation, we can estimate an arbitrary quantum observable $A(t)$, which commutes with the observation process $Y(t)$. That means $A(t)\in \mathscr{Y}_t^\prime$ \citep{LectureMatt, bouten2007introduction, dong2010quantum}. \\
\\
A typical quantum scenario in quantum optics demonstrating quantum filtering theory is a collection of atoms interacting with an electromagnetic field that is assumed to be in a vacuum state. The quantum dynamics of the atomic system is described by the following quantum stochastic differential equation \citep{bouten2007introduction, LectureMatt}:
\begin{equation}\label{eq0.03}
dU_t = \{ LdA_t^* - L^* dA_t - \frac{1}{2}L^*Ldt -iHdt \}, \quad U_0 = I,
\end{equation}
which is driven by the noncommuting white-noise process $A_t$ and $A_t^*$. The evolution of a system observable $X$ is : $X \rightarrow U^*(t)(X\otimes I)U(t)$. Also $X(t)$, which is denoted by $j_t(X)$, satisfies:
\begin{equation}\label{eq0.04}
dj_t(X)= j_t (\mathscr{L}_{L,H}(X))dt + j_t ([L^*,X])dA_t + j_t ([X,L])dA_t^*,
\end{equation}
where $\mathscr{L}$ is the quantum Lindblad generator \citep{bouten2007introduction} such that
\begin{equation}
\mathscr{L}_{L,H}(X)=i[H,X] + L^*XL - \frac{1}{2}(L^*LX+XL^*L).
\end{equation}
\\
There are two main types of measurement in quantum optics: homodyne detection and photon counting measurement. In our case, we adopt the homodyne detection scheme. The dynamic equation of the observation is
\begin{equation}\label{eq0.05}
dY_t= j_t(L+L^*)dt +dA_t +dA^*_t.
\end{equation}
Quantum filtering theory aims to provide an optimal estimate of any system observable using the observation process. From Section 2.1, this can be achieved if one can calculate the recursive equation satisfied by the conditional expectation $\pi_t(X)=\mathbb{P}(j_t(X)|\mathscr{Y}_t)$. This recursive quantum stochastic equation is then the quantum filter we obtain.
Using the reference probability method or the characteristic function method, one has \citep{bouten2007introduction}
\begin{equation}\label{eq0.06}
\begin{split}
d\pi_t(X)& =  \pi_t (\mathscr{L}_{L,H}(X))dt +    \\
& (\pi_t(L^*X+XL)-\pi_t(L^* + L)\pi_t(X))(dY_t-\pi_t(L^* +L)dt),
\end{split}
\end{equation}
or its SME form
\begin{equation}\label{eq0.07}
\begin{split}
d\rho_t = & -i[H,\rho_t]dt + (L\rho_tL^* - \frac{1}{2}L^*L\rho_t - \frac{1}{2}\rho_tL^*L)dt + \\
&(L\rho_t + \rho_t L^* -\text{Tr}[(L+L^*)\rho_t]\rho_t)dW_t,
\end{split}
\end{equation}
where the stochastic process $dW_t=dY(t)-\text{Tr}[(L+L^*)\rho_t]dt$ is a standard Wiener process. Equations \eqref{eq0.06} and \eqref{eq0.07} are quantum filter equations for open systems whose dynamics can be described by \eqref{eq0.04} and \eqref{eq0.05}.


\section{Description of Hybrid Quantum-Classical System}\label{Sec3}
In this paper, we consider a quantum cavity system disturbed by a classical diffusion stochastic process; see the schematic in Fig. 1. The classical disturbance process is assumed to evolve according to the following first-order linear stochastic differential equation (SDE)
\begin{equation}\label{eq0}
d q	= -uq dt -vdw_t,
\end{equation}
where $w_t$ is classical Wiener process with zero mean and unit variance; $u$ and $v$ are arbitrary real numbers while $u$ is assumed to be positive. The disturbance signal will influence the cavity system $S_1$ by changing its Hamiltonian such that
\begin{equation}\label{eq0.1}
H_1(t) =q(t)a^\dag(t) a(t) ,
\end{equation}
where $a(t)$ is the annihilation operator of cavity system $S_1$.
\begin{figure}[htbp]	
	\centering		
	\includegraphics[width=8.4cm]{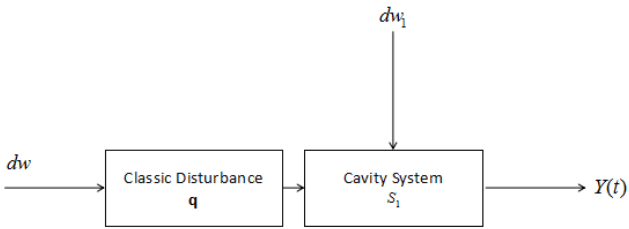}		
	\caption{A quantum cavity system $S_1$ which is affected by a classical disturbance $q(t)$. } \label{fig 1 :model1}		
\end{figure}
\\
In our case, we consider a cavity mode disturbed by an external signal \citep{gardiner1991quantum}. The dynamics of this hybrid system is different from that in the standard quantum filter problem. Later we will show how to transform the problem into a standard quantum filtering problem so that the filtering equations \eqref{eq0.06} and \eqref{eq0.07} apply. Rather than using the hybrid quantum-classical density operator method in \citep{diosi2000quantum} and \citep{aleksandrov1981statistical}, we build a quantum analog of the classical stochastic process and use quantum probability theory to analyze the combined quantum system consisting of two quantum subsystems. To be specific, we consider a cavity system with a quantum disturbance as in Fig. \ref{fig 2 :model2}, where the quantum disturbance system $S_2$ is used to model the classical disturbance signal $q$.
\begin{figure}[htbp]	
	\centering		
	\includegraphics[width=8.4cm]{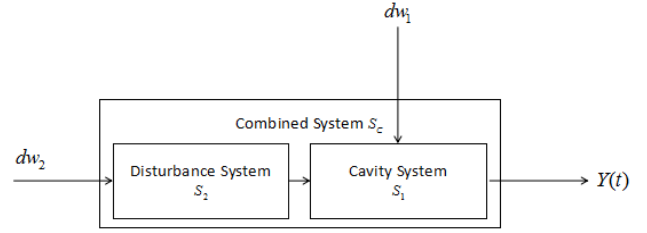}		
	\caption{A quantum cavity system $S_1$ is coupled to another quantum disturbance system $S_2$. The combined system $S_c$ consists of $S_1$ and $S_2$.} \label{fig 2 :model2}		
\end{figure}
The corresponding analogy quantum signal with respect to $q$ is $Q_2=\frac{b +b^\dag}{2}$, which is a real quadrature of the system $S_2$. That is, we write
\begin{equation}\label{0.009}
q \sim \frac{Q_2}{\alpha},
\end{equation}
where $\alpha$ is a scalar that depends on the dynamic equations of $q$ and $Q_2$. Then we can obtain an estimate of $q$ by using the relationship that
\begin{equation}\label{0.0099}
\hat{q}=\frac{\pi_t(Q_2)}{\alpha}=\frac{\text{Tr}[Q_2\hat{\rho}_2]}{\alpha},
\end{equation}
where $\pi_t(Q_2)=\mathbb{P}(Q_2|\mathscr{Y}_t)$ represents the quantum estimate of $Q_2$ given the measurement $\mathscr{Y}_t$.\\
\\
We assume that the Hamiltonian of system $S_2$ and system $S_1$ are $H_2 = 0$ and $H_1 =\frac{Q_2a^\dag a}{\alpha} $, respectively. The coupling operators are $L_2 = \sqrt{K_2} b$ and $L_1 = \sqrt{K_1} a $. $K_1,K_2 > 0$ are parameters which indicate the coupling strength to each channel.

Open quantum systems with multiple field channels can be characterized by the parameter list
\begin{equation}
G=(S,L,H)
\end{equation}
where $S$ is a scattering matrix which satisfies  $S^\dag S=I$, $L$ is the coupling vector that specifies the interface between the system and the fields and $H$ is the Hamiltonian of the system.


For the combined system $S_c$, the $(S,L,H)$ model  \citep{gough2009series} is given as
\begin{equation}\label{eq1}
\begin{split}
S &= I, \\
H &= H_1 + H_2 =\frac{Q_2}{\alpha}a^\dag a, \\
L &= \left( {\begin{array}{*{20}{c}}
	{{\sqrt{K_1} a}}   \\
	{{\sqrt{K_2} b}} \\	 	 	
	\end{array}} \right).\\
\end{split}
\end{equation}
We have now obtained a quantum system $S_2$ as the analogue of the classical disturbance system. The classical signal $q$ is now equivalently represented by $\frac{Q_2}{\alpha}=\frac{b +b^\dag}{2\alpha}$. We can derive the stochastic properties of $q$ as we obtain an estimate of $Q_2$. Moreover, we have obtained a model for the combined system $S_c$ consisting of subsystems $S_1$ and $S_2$.

Note that
\begin{equation}\label{eq2}
\begin{split}
[b,H]&=	[b, \frac{b+b^\dag}{2\alpha}a^\dag a]	\\
&=    [b, \frac{b+b^\dag}{2\alpha}]a^\dag a  \\
&=    [b, b^\dag]\frac{a^\dag a}{2\alpha}    \\
&=    \frac{a^\dag a}{2\alpha},
\end{split}
\end{equation}	 	
and that
\begin{equation}\label{eq3}
\begin{split}
[a,H]&=	[a, \frac{b+b^\dag}{2\alpha}a^\dag a]	\\
&=    \frac{b+b^\dag}{2\alpha}[a, a^\dag a]  \\
&=    \frac{b+b^\dag}{2\alpha} a    \\
&=    \frac{Q_b a}{\alpha}.
\end{split}
\end{equation}	
Then the QSDEs used to describe the disturbance system $S_2$ and cavity system $S_1$ are listed below:
\begin{equation}\label{eq4}
\begin{split}
db       &= - \frac{K_2}{2} bdt - i\frac{a^\dag a}{2\alpha} dt - \sqrt{K_2}d w_2 , \\
d b^\dag &= - \frac{K_2}{2} b^\dag dt + i\frac{a^\dag a}{2\alpha} dt - \sqrt{K_2}d w_2^\dag , \\
da       &= - \frac{K_1}{2} adt - \frac{iQ_2a}{\alpha} dt - \sqrt{K_1}d w_1 ,\\
d a^\dag &= - \frac{K_1}{2} a^\dag dt + \frac{ia^\dag Q_2}{\alpha} dt - \sqrt{K_1}d w_1^\dag . \\	 			
\end{split}
\end{equation}	
The real quadratures corresponding to the position and momentum of the two systems, respectively, are:
\begin{equation}\label{eq5}
\begin{split}
Q_1=\frac{a+a^\dag}{2} , \qquad P_1=\frac{a-a^\dag}{2i} , \qquad	[Q_1,P_1]=\frac{-i}{2} ,\\	
Q_2=\frac{b+b^\dag}{2} , \qquad P_2=\frac{b-b^\dag}{2i} , \qquad   [Q_2,P_2]=\frac{-i}{2} .
\end{split}
\end{equation}	
From
\begin{equation} \nonumber
\begin{split}
& Q_2   \sim \alpha q , \\
& dQ_2 = - \frac{K_2}{2} Q_2 dt - \frac{\sqrt{K_2}}{2}(dw_2+dw_2^*) , \\
& dq   = -uqdt-vdw_t , \\
\end{split}
\end{equation}
we have
\begin{equation} \label{0.0009}
\alpha = \frac{\sqrt{2u}}{2v} , \qquad K_2=2u =4(\alpha v)^2 .
\end{equation}
The fact $K_2 > 0$ results in $u>0$.\\
\\
A vector of operators is defined to describe the combined system $S_c$ :    	  	
\begin{equation}\label{eq6}
x = \begin{pmatrix} x_1\\ x_2\\x_3\\x_4 \end{pmatrix}  =
\begin{pmatrix} Q_1\\P_1\\Q_2\\P_2 \end{pmatrix} .		 	
\end{equation} 	
The QSDE satisfied by $x$ is \\
\begin{equation}\label{eq8}
d x	= f(x) dt+G dz_x + G^* d z_x^* ,
\end{equation}
where
\begin{equation}\label{eq7}
\begin{split}
z_x&=	\begin{pmatrix} w_1\\  w_2	\end{pmatrix} ,\qquad \qquad \qquad z_x^*=	\begin{pmatrix}  w_1^\dag\\ w_2^\dag	\end{pmatrix} 	 ;\\
G&=  \begin{pmatrix} -\frac{\sqrt{K_1}}{2} & 0 \\  -\frac{\sqrt{K_1}}{2i}  & 0\\  0 &-\frac{\sqrt{K_2}}{2} \\  0 &-\frac{\sqrt{K_2}}{2i}	 \end{pmatrix} , \quad
G^* =   \begin{pmatrix} -\frac{\sqrt{K_1}}{2} & 0 \\  -i \frac{\sqrt{K_1}}{2} & 0\\ 0  &-\frac{\sqrt{K_2}}{2} \\  0 &-i \frac{\sqrt{K_2}}{2}	 \end{pmatrix}; \\
f(x)&= \begin{pmatrix} -\frac{K_1}{2}x_1+ \frac{x_2x_3}{\alpha}\\-\frac{K_1}{2} x_2-\frac{x_1x_3}{\alpha}\\-\frac{K_2}{2}x_3\\-\frac{K_2}{2}x_4-\frac{1}{2\alpha}x_1^2-\frac{1}{2\alpha}x_2^2-\frac{1}{4\alpha} \end{pmatrix} .
\end{split}
\end{equation}
The homodyne detection method is used to continuously monitor the scattered field from the cavity $S_1$, which generates an observation process satisfying
\begin{equation}\label{eq10}
d y = (L_1 + L_1^\dag )dt + dw_y +dw_y^* .		 	
\end{equation}
Let $C = (2 \sqrt{K_1} \quad 0 \quad 0 \quad 0) $, $h(x)=Cx$ and $dz_y=dw_y +dw_y^*$.  Equation \eqref{eq10} can be rewritten in the following compact form
\begin{equation}\label{eq11}
d y = h(x) dt + dz_y.		 	
\end{equation} 	
Then the evolution of the system $S_c$ in the Heisenberg picture can be described by the diffusive QSDEs \eqref{eq8} and \eqref{eq11}. Since the combined system consists of two quantum subsystems, quantum filtering theory can be directly applied to the combined system and the standard quantum filter is described by the stochastic master equation (SME)\\
\begin{equation}\label{eq18}
\begin{split}
d\rho_t & =  -i[H,\rho_t]dt + (L\rho_tL^* - \frac{1}{2}L^*L\rho_t - \frac{1}{2}\rho_tL^*L)dt + \\
&(L\rho_t + \rho_t L^* -\text{Tr}[(L+L^*)\rho_t]\rho_t)(dY(t)-\text{Tr}[(L+L^*)\rho_t]dt),
\end{split}
\end{equation}
where the corresponding $H$ and $L$ are given in (\ref{eq1}).\\
\section{Extended Kalman Filter}\label{Sec4}
The fact that the computation time in simulating the filter SME scales exponentially with the dimension of the Hilbert space adds difficulty to implementing the filter in realtime. A quantum extended Kalman filter (QEKF) was introduced in \citep{emzir2016quantum}, aiming to reduce the computational complexity of the quantum filter. For the QEKF, the constraint that elements of observable operator vector $x(t)$ belong to a commutative von Neumann algebra is not required  and there can be non-commutating operators in the dynamic equation. Otherwise, if the commutativity of all the observables and operators are given, then the QEKF reduces to a classical EKF since the filtering problem can be transformed into a single classical probability space using the $*$-isomorphism. This is the main difference between the classical EKF method and the QEKF method. A commutative operator approximation of the non-commutative nonlinear QSDE is used to estimate the system observables given nondemolition measurements. Keeping the first order term of the Taylor series, the Kalman filter gain is effectively calculated. This method was proposed to solve the filtering problem for a class of multiple output channel open quantum systems whose evolution can be described by the following QSDE \citep{emzir2016quantum}:
\begin{equation} \label{eq12}
dx_t = f(x_t)dt + G(x_t)dA_t^* + G(x_t)^*dA_t ,
\end{equation}
where
\begin{equation} \label{eq13}
f(x_t)=\mathscr{L}(x_t) \quad \text{and} \quad G(x_t) = [x_t,\mathbb{L}_t]\mathbb{S}_t^* .
\end{equation}
The operators $\mathbb{L}$ and $\mathbb{S}$ are the parameters from the $(\mathbb{S},\mathbb{H},\mathbb{L})$ model which can be used to describe a multiple channel open quantum system \citep{emzir2016quantum}.
The measurement dynamic equation is given by
\begin{equation} \label{eq14}
dy_t = h(x_t)dt + L(x_t)dA_t^* + L(x_t)^*dA_t +N_td\alpha_t ,
\end{equation}
with
\begin{equation} \label{eq15}
\begin{split}
&h(x_t)=E_t^*\mathbb{L}+E_t\mathbb{L}^* + N_t1_t;\\
&L(x_t)=(E_t+N_t\overline{\mathbb{L}})\mathbb{S}_t^*  ,
\end{split}
\end{equation}
where $\overline{\mathbb{L}} =\mathbb{L}= L$ and $1_t=L^*L$ in our case. $E_t$ and $N_t$ are real matrices. $d\alpha_t = \text{diag}(\mathbb{S}_t d \Lambda \mathbb{S}_t^\top)$ where $\Lambda$ is the conservation process that represents the photon counting measurement.  $E_t$ indicates output channels which are subject to the homodyne detection measurement. $N_t$ shows photon counting measurement channels. For example, if a quantum system is observed by a homodyne detector and a photon counting measurement, we have
\begin{equation}
E=\begin{pmatrix} 1 \quad 0\\  0 \quad 0	\end{pmatrix} ,	 \quad N=\begin{pmatrix} 0\quad 0\\ 0 \quad 1	\end{pmatrix} .
\end{equation}
According to \citep{emzir2016quantum}, $E_t$ and $N_t$ have to satisfy the condition of Theorem 3.1 in \citep{emzir2015quantum}. For this paper, this condition is satisfied since we have $E=1$ and $N=0$ which means the system is observed using only one homodyne detector.  \\
\\
Let $\mathcal{C}_{op}^1(I)$ denote the Banach $*$-algebra of $\mathcal{C}^1$-functions on the compact interval $I$ such that the corresponding Hilbert space operator function $T \rightarrow f(T)$, for $T=T^*$ and the spectra of $T$ satisfies $\text{sp}(T)\in I$, is Fr\'echet differentiable \citep{pedersen2000operator}. The Fr\'echet derivative of an operator differentiable function $f \in \mathcal{C}_{op}^1 (I)$ can then be constructed as in the following lemma:
 \begin{lem} \label{prop3}
 \citep{pedersen2000operator} If $f \in \mathcal{C}_{op}^1 (I)$, for any two elements $S,T \in \mathcal{U}$, a unital commutative $\mathcal{C}^*$- algebra, then the corresponding Fr\'echet derivative satisfies
 \begin{equation}
 D_{(f,T)}S = f^\prime (T)S,
 \end{equation}
 \end{lem}
where $D_{(f,T)}$ is the Fr\'echet derivative and $f^\prime (\cdot)$ denotes the normed derivative of $f(\cdot)$. \\
\\
This lemma can be used to calculate the partial derivative of the nonlinear quantum Markovian process generator $f(x)$ in case $f(x)\in \mathcal{C}_{op}^1 (I)$. \\
\\
According to \citep{pedersen2000operator}, we have $\mathcal{C}^2 \subseteq \mathcal{C}_{op}^1 \subseteq \mathcal{C}^1$ which means if the function $f \in \mathcal{C}^2$, then its operator extension is operator differentiable.  Given \eqref{eq1} and let $f(x_t)= \mathscr{L}_{H,L}(x_t)$, we have $f(x_t) \in \mathcal{C}^2(R)$ which results in $f(x_t) \in \mathcal{C}_{op}^1 (R)$.

Note that, using the QEKF method, $\hat{x}_t$ is no longer the projection of $x_t$ onto $\mathscr{Y}_t$. That is, $\hat{x}_t \ne \mathbb{E}_{\mathbb{P}}[x_t|\mathscr{Y}_t]$. However, if $\hat{x}_0 \in \mathscr{Y}_0$, then we still have $\hat{x}_t \in \mathscr{Y}_t$ and the elements of $\hat{x}_t$ are commutative with other elements.\\
\\
According to Lemma 3, one can calculate
\begin{equation}\label{eq22}
\begin{split}
& F(x)=f^{'}(x) =  \begin{pmatrix} -\frac{K_1}{2} & \frac{x_3}{\alpha} & \frac{x_2}{\alpha} & 0\\ \frac{-x_3}{\alpha} & -\frac{K_1}{2} & \frac{-x_1}{\alpha} & 0\\ 0 & 0 &-\frac{K_2}{2} & 0\\ \frac{-x_1}{\alpha} & \frac{-x_2}{\alpha} & 0 &-\frac{K_2}{2}	\end{pmatrix} ,  \\
& H(x)=h^{'}(x) = \begin{pmatrix} 2\sqrt{K_1} & 0 & 0 & 0\end{pmatrix} .
\end{split}		 	
\end{equation}
\\
As in \citep{emzir2016quantum}, let the variance of the system observables and measurements be denoted as $Q_t$ and $R_t$, respectively. Also, the cross-correlation matrix of the system observables and measurements is denoted as $S_t$, such that
\begin{equation}\label{eq23}
\begin{split}
Q_t&=\frac{1}{2dt}\mathbb{E}_\mathbb{P}[\{dx_t,dx_t\}|\mathscr{Y}_t] ;\\
R_t&=\frac{1}{2dt}\mathbb{E}_\mathbb{P}[\{dy_t,dy_t\}|\mathscr{Y}_t] ;\\
S_t&=\frac{1}{2dt}\mathbb{E}_\mathbb{P}[\{dx_t,dy_t\}|\mathscr{Y}_t] .\\
\end{split}		 	
\end{equation}
The anti-commutator above is given by $\{x,y\}=xy^\top + (yx^\top)^\top$. In our case of the combined system $S_c$ , \eqref{eq23} yields
\begin{equation}\label{eq24}
\begin{split}
Q_t&=\frac{1}{2dt}\mathbb{E}_\mathbb{P}[(GG^\dag + (GG^\dag)^\top) dt|\mathscr{Y}_t] = \frac{1}{2}(GG^\dag + (GG^\dag)^\top) ;\\
R_t&=\frac{1}{2dt}\mathbb{E}_\mathbb{P}[2 Idt|\mathscr{Y}_t] = I ;\\
S_t&=0 .\\
\end{split}		 	
\end{equation}
\\
To apply the QEKF in our case, the following constraints should be satisfied:
\begin{enumerate}
	\item [(i)] The covariance and cross-correlation matrices $Q_t$, $R_t$, $S_t$ are single valued (see Definition 2.4 in \citep{emzir2016quantum} for the definition of a single valued operator);
	\item [(ii)] $R_t$ is invertible;
	\item [(iii)] Initially $\hat{x}_0 \in \mathscr{Y}_0$ .
\end{enumerate}
It can be verified that the first two constraints are satisfied in our case. As a result, we only have to make sure that $\hat{x}_0 \in \mathscr{Y}_0$.
Then the quantum EKF can be given as
\begin{equation}\label{eq27}
d\hat{x}_t = f(\hat{x}_t) dt + K_t (dy_t - d\hat{y}_t) ,
\end{equation}
where $P_t$ is defined as $P_t\triangleq \frac{1}{2}\mathbb{E}_{\mathbb{P}}[\{\tilde{x}_t,\tilde{x}_t\}|\mathscr{Y}_t]$ and $\tilde{x}_t=x_t - \hat{x}_t$. $P_t$ evolves according to the following Riccati differential equation \citep{emzir2016quantum}
\begin{equation}\label{eq26}
\begin{split}
\frac{dP_t}{dt} = & F(\hat{x}_t)P_t + P_t	F(\hat{x}_t)^\top + \\
                  & Q_t - [P_t	H(\hat{x}_t)^\top + S_t] 	 R_t^{-1} 	[P_t	H(\hat{x}_t)^\top + S_t]^\top .
\end{split}
\end{equation}
Without loss of generality, we assume that $P_0 \in \mathscr{Y}_0$.
Consider an open quantum system described by the QSDEs given in \eqref{eq8} subject to the measurements given in \eqref{eq10}. From the result in \citep{emzir2016quantum}, then there exists a Kalman gain $K_t \in \mathscr{Y}_t$,
\begin{equation}\label{eq25}
K_t = [P_t H(\hat{x}_t) + S_t]R_t^{-1} ,		 	
\end{equation}
such that if the quantum extended Kalman filter is given by \eqref{eq27}, then $\hat{x}_t \in \mathscr{Y}_t , \forall t\geqslant 0$ and $P_t$ evolves according to \eqref{eq26} upon neglecting the residual terms of the Taylor series.

To implement numerical calculations using the QEKF method, one needs to transform \eqref{eq27} into a classical stochastic differential equation. This is feasible since we are only concerned with the mean value and covariance of $x(t)$ for our application. Recall that $\mathscr{Y}_t$ is a commutative von Neumann algebra generated by the measurement $dy_t$. By Theorem 2, there exists a $*$-isomorphism $\iota$ from $\mathscr{Y}_t$ to $L^\infty(\Omega, \mathcal{F},\mu)$. Denoting $\iota(\cdot)_{t,w}$ as $(\cdot)_{t,w}$, the following classical SDE is satisfied \citep{emzir2016quantum}:
\begin{equation}\label{eq31}
d\hat{x}_{t,w} = [f(\hat{x}_{t,w}) - K_{t,w}h(\hat{x}_{t,w})]dt + K_{t,w}dy_{t,w},
\end{equation}
for all $w\in\Omega, t\geq 0$. The dynamic equation \eqref{eq26} for $P_t$ can also be written as a classical SDE in the same way. Since $\hat{x}_{t,w}$, $dy_{t,w}$ and $P_{t,w}$ are classical random variables in the same probability space $L^\infty(\Omega, \mathcal{F},\mu)$, then the previous constraints $\hat{x}_0 \in \mathscr{Y}_0$ and $P_0 \in \mathscr{Y}_0$ can all be satisfied (for details, see \citep{emzir2016quantum}). The QEKF method is thus suitable for our problem.   \\
\section{Numerical example}\label{Secsim}
\begin{figure*}	
	\centering		
	\includegraphics[width=13cm]{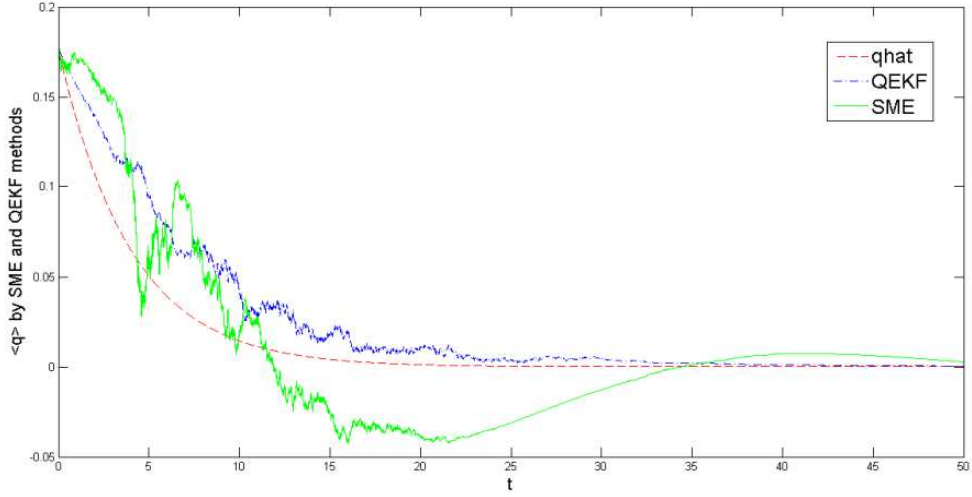}		
	\caption{ Application of the quantum EKF and SME methods to estimate the real quadrature $\frac{Q_2}{\alpha}$. The red line is the ensemble average (qhat) of 500 trajectories of the classical stochastic process $q$. The green line is the estimate of quantum real quadrature $\frac{Q_2}{\alpha}$ using the SME method. The blue line is the estimate of $\frac{Q_2}{\alpha}$ as calculated by the QEKF method.} \label{fig 3 :test1}		
\end{figure*}
\begin{figure*}
	\centering		
	\includegraphics[width=13cm]{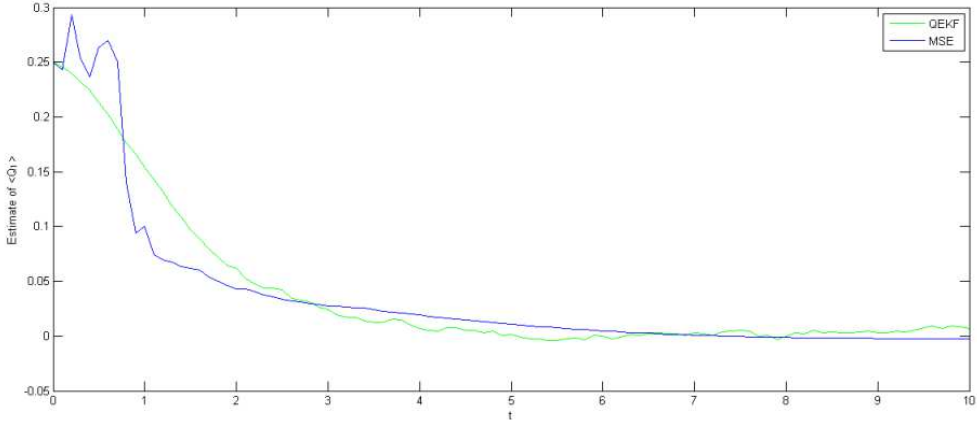}		
	\caption{ Application of the quantum EKF and SME methods to estimate the real quadrature $Q_1$.  The blue line is the estimate of quantum real quadrature $Q_1$ using the SME method. The green line is the estimate of $Q_1$ as calculated by the QEKF method.} \label{fig 5 :test1}		
\end{figure*}
\begin{figure*}	
	\centering		
	\includegraphics[width=13cm]{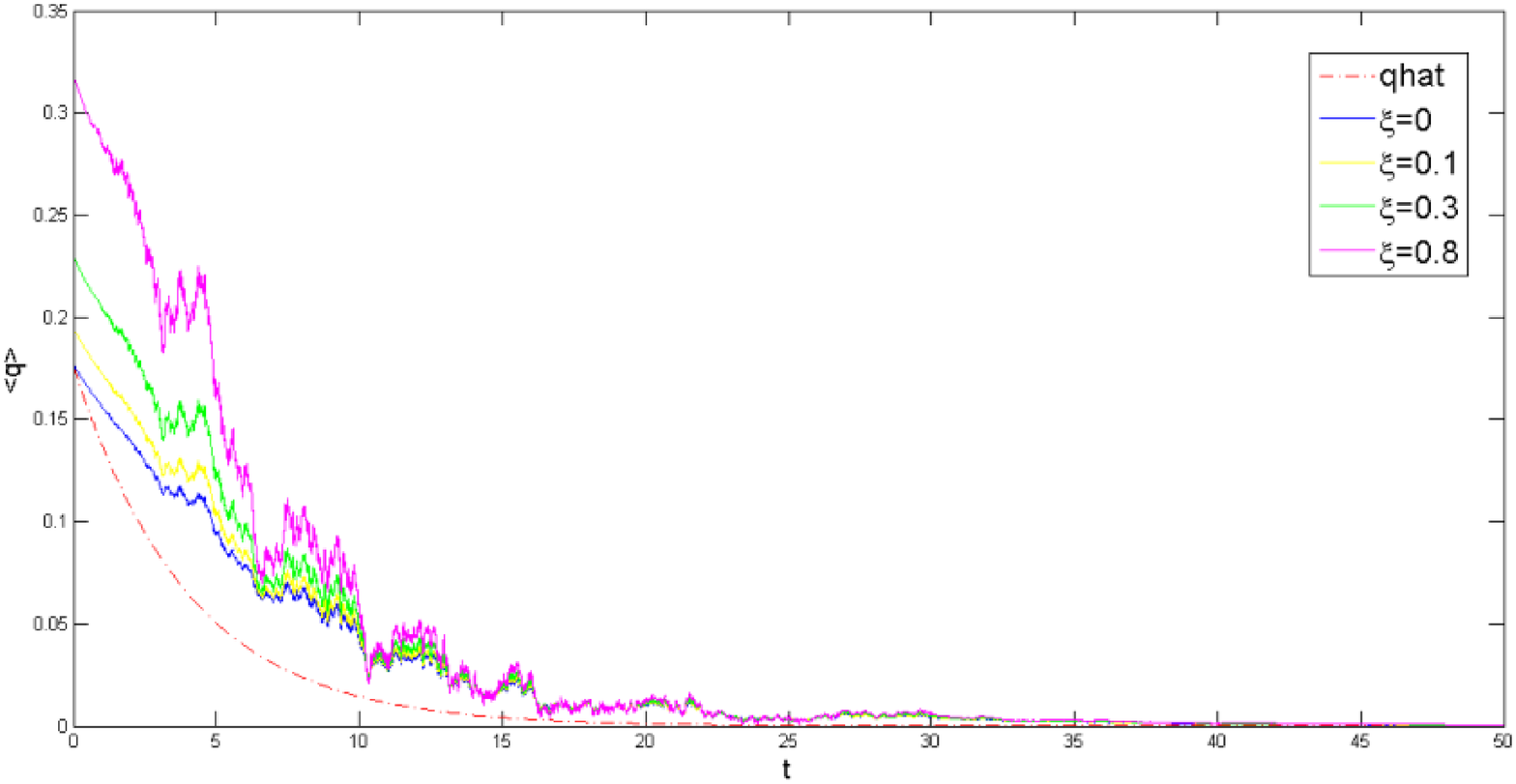}		
	\caption{The estimate of $q$ using the QEKF method with different initial errors. The error matrix is set as $\hat{X}_0=\hat{X}_{SME} + \xi [0.5 \quad 0.5 \quad 0.5 \quad 0.5]^\top$.} \label{fig 4 :test2}		
\end{figure*}

In our example, the evolution of both the system $S_1$ and the system $S_2$ can be represented by the annihilators $a(t)$ and $b(t)$. The aim is to estimate the real quadrature $Q_2=\frac{b+b^*}{2}$ of system $S_2$. Then we can obtain an estimate of both the quantum quadrature $Q_a$ and the classical signal $q$ using the relationship $\hat{q}=\frac{\pi_t(Q_2)}{\alpha}=\frac{\text{Tr}[Q_2\hat{\rho}_b]}{\alpha}$ . \\
\\
The basic settings are listed below:
\begin{equation}
dq = - \frac{1}{4}q dt  -\frac{1}{8} dw_t ,
\end{equation}
which means that $\alpha= - \frac{\sqrt{u}}{\sqrt{2}v}=2\sqrt{2}$ and $K_2=2u=0.5$. The initial quantum states for $S_1$ and $S_2$ are $\rho_1=\rho_2=\frac{1}{2}(I + \sigma_x)$. We also choose $K_1=0.55$. The initial value of $q$ is set to be $\frac{1}{4\sqrt{2}}$. Fig. 3 illustrates the trajectories of $\frac{\hat{Q}_2}{\alpha}$ obtained by using the SME and QEKF methods, respectively. In the SME method, each cavity is approximated by a two level system. The red line is the average value of $q$ over 500 realizations. It can be seen that the estimated $\hat{q}=\frac{\hat{Q}_2}{\alpha}$ for both methods converges to the real expected value of $q$.  Fig. 4 demonstrate the estimate of quantum real quadrature $Q_1$ using the SME method and the QEKF method respectively.

In order to test the robustness of our method, a set of perturbations on the initial state is considered in Fig. 5. A certain level of initial error can affect the performance of the QEKF method but the convergence is still guaranteed.


\section{Conclusion}\label{Conclusion}

 By modeling a classical stochastic process using a quantum cavity model, we solve the filtering problem of a class of hybrid quantum-classical systems using the standard quantum filtering method and a quantum extended Kalman filtering method. A performance comparison between these two methods is provided using numerical results. Future work includes extending our method to more general quantum systems and classical signals with nonlinear dynamics.\\

\begin{ack}
Discussions with Muhammad Fuady Emzir are gratefully acknowledged.
\end{ack}

\bibliography{ifacconf}             

\begin{thebibliography}{30}
\providecommand{\natexlab}[1]{#1}
\providecommand{\url}[1]{\texttt{#1}}
\providecommand{\urlprefix}{URL }
\expandafter\ifx\csname urlstyle\endcsname\relax
  \providecommand{\doi}[1]{doi:\discretionary{}{}{}#1}\else
  \providecommand{\doi}{doi:\discretionary{}{}{}\begingroup
  \urlstyle{rm}\Url}\fi

\bibitem[{Aleksandrov(1981)}]{aleksandrov1981statistical}
Aleksandrov, I. (1981).
\newblock The statistical dynamics of a system consisting of a classical and a
  quantum subsystem.
\newblock \emph{Zeitschrift f{\"u}r Naturforschung A}, 36(8), 902--908.

\bibitem[{Armen et~al.(2002)Armen, Au, Stockton et~al.}]{armen2002adaptive}
Armen, M.A., Au, J.K., Stockton, J.K., et~al. (2002).
\newblock Adaptive homodyne measurement of optical phase.
\newblock \emph{Physical Review Letters}, 89(13), 133602.

\bibitem[{Belavkin(1991)}]{belavkin1991continuous}
Belavkin, V. (1991).
\newblock Continuous non-demolition observation quantum filtering and optimal
  estimation.
\newblock In \emph{Quantum aspects of optical communications}, 151--163.
  Springer.

\bibitem[{Bouten et~al.(2007)Bouten, van Handel, and
  James}]{bouten2007introduction}
Bouten, L., van Handel, R., and James, M.R. (2007).
\newblock An introduction to quantum filtering.
\newblock \emph{SIAM Journal on Control and Optimization}, 46(6), 2199--2241.

\bibitem[{Bouten et~al.(2009)Bouten, Van~Handel, and
  James}]{bouten2009introduction}
Bouten, L., Van~Handel, R., and James, M.R. (2009).
\newblock A discrete invitation to quantum filtering and feedback control.
\newblock \emph{SIAM review}, 51(2), 239--316.

\bibitem[{Di{\'o}si et~al.(2000)Di{\'o}si, Gisin, and
  Strunz}]{diosi2000quantum}
Di{\'o}si, L., Gisin, N., and Strunz, W.T. (2000).
\newblock Quantum approach to coupling classical and quantum dynamics.
\newblock \emph{Physical Review A}, 61(2), 022108.

\bibitem[{Dong and Petersen(2010)}]{dong2010quantum}
Dong, D. and Petersen, I.R. (2010).
\newblock Quantum control theory and applications: a survey.
\newblock \emph{IET Control Theory \& Applications}, 4(12), 2651--2671.

\bibitem[{Emzir et~al.(2015)Emzir, Woolley, and Petersen}]{emzir2015quantum}
Emzir, M.F., Woolley, M.J., and Petersen, I.R. (2015).
\newblock Quantum filtering for multiple diffusive and poissonian measurements.
\newblock \emph{Journal of Physics A: Mathematical and Theoretical}, 48(38),
  385302.

\bibitem[{Emzir et~al.(2016)Emzir, Woolley, and Petersen}]{emzir2016quantum}
Emzir, M.F., Woolley, M.J., and Petersen, I.R. (2016).
\newblock A quantum extended kalman filter.
\newblock \emph{arXiv preprint arXiv:1603.01890}.

\bibitem[{Gao et~al.(2016{\natexlab{a}})Gao, Dong, and Petersen}]{gao2016fault}
Gao, Q., Dong, D., and Petersen, I.R. (2016{\natexlab{a}}).
\newblock Fault tolerant quantum filtering and fault detection for quantum
  systems.
\newblock \emph{Automatica}, 71, 125--134.

\bibitem[{Gao et~al.(2016{\natexlab{b}})Gao, Dong, Petersen
  et~al.}]{gao2016fault2}
Gao, Q., Dong, D., Petersen, I.R., et~al. (2016{\natexlab{b}}).
\newblock Fault tolerant filtering and fault detection for quantum systems
  driven by fields in single photon states.
\newblock \emph{Journal of Mathematical Physics}, 57(6), 062201.

\bibitem[{Gardiner and Haken(1991)}]{gardiner1991quantum}
Gardiner, C.W. and Haken, H. (1991).
\newblock \emph{Quantum noise}, volume~26.
\newblock Springer Berlin.

\bibitem[{Gough and James(2009)}]{gough2009series}
Gough, J. and James, M.R. (2009).
\newblock The series product and its application to quantum feedforward and
  feedback networks.
\newblock \emph{IEEE Transactions on Automatic Control}, 54(11), 2530--2544.

\bibitem[{Hou et~al.(2016)Hou, Zhong, Tian, Dong, Qi, Li, Wang, Nori, Xiang,
  Li, and Guo}]{hou:2016}
Hou, Z., Zhong, H.S., Tian, Y., Dong, D., Qi, B., Li, L., Wang, Y., Nori, F.,
  Xiang, G.Y., Li, C.F., and Guo, G.C. (2016).
\newblock Full reconstruction of a 14-qubit state within four hours.
\newblock \emph{New Journal of Physics}, 18(8), 083036.

\bibitem[{Hume et~al.(2007)Hume, Rosenband, and Wineland}]{hume2007high}
Hume, D., Rosenband, T., and Wineland, D. (2007).
\newblock High-fidelity adaptive qubit detection through repetitive quantum
  nondemolition measurements.
\newblock \emph{Physical Review Letters}, 99(12), 120502.

\bibitem[{James(2015)}]{LectureMatt}
James, M.R. (2015).
\newblock Quantum measurement, lecture notes phys4003b.
\newblock The Australian National University.

\bibitem[{Paris and \v{R}eh\'{a}\v{c}ek(2004)}]{book1}
Paris, M. and \v{R}eh\'{a}\v{c}ek, J.e. (2004).
\newblock \emph{Quantum State Estimation}, volume 649 of Lecture Notes in
  Physics.
\newblock Springer Berlin.

\bibitem[{Pedersen(2000)}]{pedersen2000operator}
Pedersen, G.K. (2000).
\newblock Operator differentiable functions.
\newblock \emph{Publications of the Research Institute for Mathematical
  Sciences}, 36(1), 139--157.

\bibitem[{Qi et~al.(2013)Qi, Hou, Li, Dong, Xiang, and Guo}]{qi:2013}
Qi, B., Hou, Z., Li, L., Dong, D., Xiang, G.Y., and Guo, G.C. (2013).
\newblock Quantum state tomography via linear regression estimation.
\newblock \emph{Sci. Rep.}, (3), 3496.

\bibitem[{Ralph et~al.(2011)Ralph, Jacobs, and Hill}]{ralph2011frequency}
Ralph, J.F., Jacobs, K., and Hill, C.D. (2011).
\newblock Frequency tracking and parameter estimation for robust quantum state
  estimation.
\newblock \emph{Physical Review A}, 84(5), 052119.

\bibitem[{Sayrin et~al.(2011)Sayrin, Dotsenko, Zhou et~al.}]{sayrin2011real}
Sayrin, C., Dotsenko, I., Zhou, X., et~al. (2011).
\newblock Real-time quantum feedback prepares and stabilizes photon number
  states.
\newblock \emph{Nature}, 477(7362), 73--77.

\bibitem[{Tsang(2009{\natexlab{a}})}]{tsang2009optimal}
Tsang, M. (2009{\natexlab{a}}).
\newblock Optimal waveform estimation for classical and quantum systems via
  time-symmetric smoothing.
\newblock \emph{Physical Review A}, 80(3), 033840.

\bibitem[{Tsang(2009{\natexlab{b}})}]{tsang2009time}
Tsang, M. (2009{\natexlab{b}}).
\newblock Time-symmetric quantum theory of smoothing.
\newblock \emph{Physical Review Letters}, 102(25), 250403.

\bibitem[{van Handel et~al.(2005)van Handel, Stockton, and
  Mabuchi}]{van2005feedback}
van Handel, R., Stockton, J.K., and Mabuchi, H. (2005).
\newblock Feedback control of quantum state reduction.
\newblock \emph{IEEE Transactions on Automatic Control}, 50(6), 768--780.

\bibitem[{Wang and Dong(2016)}]{wang2016fault}
Wang, S. and Dong, D. (2016).
\newblock Fault-tolerant control of linear quantum stochastic systems.
\newblock \emph{IEEE Transactions on Automatic Control}, PP(99), 1--1.

\bibitem[{Wang et~al.(2013)Wang, Nurdin, Zhang, and James}]{wang2013quantum}
Wang, S., Nurdin, H.I., Zhang, G., and James, M.R. (2013).
\newblock Quantum optical realization of classical linear stochastic systems.
\newblock \emph{Automatica}, 49(10), 3090--3096.

\bibitem[{Wieczorek et~al.(2015)Wieczorek, Hofer, Hoelscher-Obermaier
  et~al.}]{wieczorek2015optimal}
Wieczorek, W., Hofer, S.G., Hoelscher-Obermaier, J., et~al. (2015).
\newblock Optimal state estimation for cavity optomechanical systems.
\newblock \emph{Physical Review Letters}, 114(22), 223601.

\bibitem[{Xue et~al.(2016)Xue, Hush, and Petersen}]{xue2016feedback}
Xue, S., Hush, M.R., and Petersen, I.R. (2016).
\newblock Feedback tracking control of non-markovian quantum systems.
\newblock \emph{IEEE Transactions on Control Systems Technology}, PP(99),
  1--12.

\bibitem[{Xue et~al.(2015{\natexlab{a}})Xue, James, Shabani, Ugrinovskii, and
  Petersen}]{xue2016quantum}
Xue, S., James, M.R., Shabani, A., Ugrinovskii, V., and Petersen, I.R.
  (2015{\natexlab{a}}).
\newblock Quantum filter for a class of non-markovian quantum systems.
\newblock In \emph{Decision and Control (CDC), 2015 IEEE 54th Annual Conference
  on}, 7096--7100. IEEE.

\bibitem[{Xue et~al.(2015{\natexlab{b}})Xue, James, Shabani, Ugrinovskii, and
  Petersen}]{xue2015quantum}
Xue, S., James, M.R., Shabani, A., Ugrinovskii, V., and Petersen, I.R.
  (2015{\natexlab{b}}).
\newblock Quantum filter for a non-markovian single qubit system.
\newblock In \emph{Control Applications (CCA), 2015 IEEE Conference on},
  19--23. IEEE.

\end{thebibliography}

\end{document}